\documentclass[10pt,twoside]{article}

\usepackage{fancyhdr}
\usepackage{titlesec}
\usepackage{cite}
\usepackage{ifthen}
\usepackage{graphicx}
\usepackage{float}
\usepackage{amsmath,amssymb}

\titleformat{\section}{\centering\large\bfseries}{\S\arabic{section}}{1em}{}
\newboolean{first}
\setboolean{first}{true}

\begin{document}

\setlength\abovedisplayskip{2pt}
\setlength\abovedisplayshortskip{0pt}
\setlength\belowdisplayskip{2pt}
\setlength\belowdisplayshortskip{0pt}

\title{\bf \Large  The robustness of multiplex networks under layer node-based attack
\author{Dawei Zhao$^{1}$   \  \ Lianhai Wang$^{1}$\ \ Zhen Wang$^{2,3}$ \\ \small \it $^{1}$Shandong Provincial Key Laboratory of Computer
Network, Shandong Computer\\
\small \it  Science Center (National Supercomputer
Center in Jinan), Jinan 250014,
China \\ \small \it $^{2}$ Interdisciplinary Graduate School of Engineering Sciences, \\ \small \it Kyushu
University, Kasuga-koen, Kasuga-shi, Fukuoka 816-8580, Japan \\ \small \it $^{3}$ School of Automation, Northwestern Polytechnical University, Xi¡¯an 710072, China}\date{}} \maketitle

\footnote{E-mail address: zhaodw@sdas.org (Dawei Zhao)}

\begin{center}
\begin{minipage}{135mm}
{\bf \small Abstract}.\hskip 2mm {\small From transportation networks to complex infrastructures, and to social and economic networks, a large variety of systems can be described in terms of multiplex networks formed by a set of nodes interacting through different network layers. Network robustness, as one of the most successful application areas of complex networks, has also attracted great interest in both theoretical and empirical researches. However, the vast majority of existing researches mainly focus on the robustness of single-layer networks an interdependent networks, how multiplex networks respond to potential attack is still short of further exploration. Here we study  the robustness of multiplex networks under two attack strategies: layer node-based
random attack and layer node-based targeted attack. A theoretical analysis framework is proposed to calculate the critical threshold and the size of giant component of  multiplex networks when a fraction of layer nodes are removed randomly or intentionally. Via numerous simulations, it is unveiled that the theoretical method can
accurately predict the threshold and the size of  giant component,
irrespective of attack strategies. Moreover, we also compare the
robustness of multiplex networks under multiplex node-based attack
and layer node-based attack, and find that layer node-based attack
makes multiplex networks more vulnerable, regardless of average
degree and underlying topology. Our finding may shed new light on
the protection of multiplex networks.}
\end{minipage}\end{center}
\begin{center}
\begin{minipage}{135mm}
{\bf \small Keyword}.\hskip 2mm {\small Multiplex network, Robustness, Layer node-based attack}
\end{minipage}
\end{center}

\section{Introduction}

Robustness of networks refers to the ability of preserving their
functional integration when they are subject to failures or attacks
\cite{Cohen10,Callaway00}. Understanding the robustness of networks
is thus useful for evaluating the resilience of systems and
constructing more efficient architectures. During the past decades,
there have been a great number of works contributing to this topic. But the majority of these achievements mainly focus on the vulnerability of single-layer networks \cite{Albert00N,Motter02PRE,Wang07EPL,Shargel03PRL,Xiao10EPL}, which
seems inconsistent with the well-recognized fact that nodes can simultaneously be the elements of more than one network in most, yet not all, natural and social systems \cite{mul1,mul2,zhenl5}. Recently, Buldyrev {\it et al.} studied the robustness of interdependent networks, where two networks were
coupled in one-to-one interdependence way \cite{Buldyrev10}.
Following the failure of one node, a cascading crash took place in
both networks (namely, interdependent networks are intrinsically
more fragile than traditional single-layer networks), which was accurately
validated by the theoretical analysis as well. After this
interesting finding, the research of network science is fast
extended to multilayer framework
\cite{Boccaletti14,Domenico13,Kivela13,Salehi14,Zhao15AMC}, where systems are
usually composed of several network layers, including interdependent
networks
\cite{Gao12NP,Dong13PRE,Gao11PRL,Wang13SR,Parshani10,Wang13,Parshani10EPL,Shao11PRE},
interconnected networks
\cite{Radicchi13NP,Dickison12PRE,Domenico14PNAS,Saumell12PRE,Wang13PRE,Zhao14PS}
and multiplex networks
\cite{Gomez13PRL,Nicosia13PRL,Sole13PRE,Kouvaris14ARXIV,Granell13PRL,Zhao14PLA,Zhao14PO,Gambuzza14,Battiston14PRE,Kim13PRL,Garde12SR,Buono14PL,Min14PRE}.
Thus far, the topological characteristics of multilayer
networks and dynamical process (such as evolutionary game theory
\cite{Wang13SR,Wang13}, disease spreading
\cite{Saumell12PRE,Dickison12PRE,Zhao14PLA,Buono14PL}, random
diffusion \cite{Gomez13PRL} and synchronization \cite{Gambuzza14})
upon them have attracted great attention in both theoretical and
empirical areas (for a recent review see \cite{Boccaletti14}).

\begin{figure}[!htb]
\centering
\includegraphics[scale=0.8,trim=50 0 50 0]{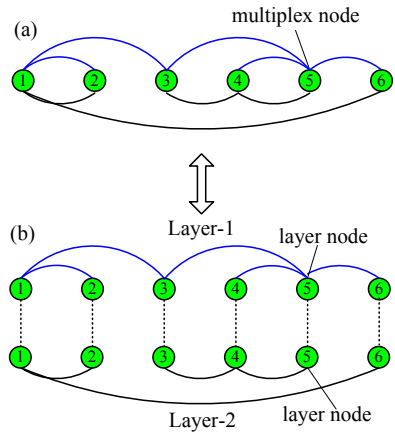}
\caption{(a) six nodes are connected via two kinds of links, blue link and black link. (b) Such systems can be embedded into the framework of multiplex networks with two types of links. Each link type defines a network layer, and the nodes of each network layer are same. The connectivity inter-layer (dash line) is from each node to itself.}
\end{figure}

Different from interdependent netowrks, multiplex networks, as a typical kind of topology structures, can be regarded as the combination of several network layers which contain the same nodes yet different intra-layer connections. In this sense, many
real-world systems like online social networks
\cite{Dodds03}, technological networks
\cite{Wang05PRL}, transportation networks
\cite{Banavar99N} can be further studied with the viewpoint of
multiplex networks. Fig.1 gives an illustration of multiplex framework: six people are connected via two kinds of relationship, for example Facebook friends (blue links) and Twitter friends (black links) (panel (a)). Such systems can be well embedded into the framework of multiplex networks with two types of links. Each link type in the system defines a network layer, and the nodes of each network layer are same (see panel (b), the connectivity inter-layer is from each node to itself). To distinguish the node of multiplex networks (nodes in panel (a))
and its agent in each network layer, here we define them the
terminology respectively: multiplex node and layer node.

Looking back to the early topic, the research of robustness of
multiplex networks thus becomes a very interesting and crucial
challenge. In \cite{Min14PRE}, Min {\it et al.} explored the
robustness of multiplex networks when multiplex nodes were removed
randomly or intentionally (here the removal of a multiplex node
means all its agents in each network layers are removed). They showed that correlated coupling would affect the structural robustness of
multiplex networks in diverse fashion. In some realistic cases, however, the failure unites or attack targets may be just the layer nodes. For example, the users of social networks are banned to use one or some but not all of the social network sites. Similarly, for multiplex transport networks where nodes are cities and network layers are airplane network, highway network and railway network, the failures may take place in one or some but not all layers. Therefore, an interesting question naturally poses itself,
which we aim to address in this letter. Namely, how does the removal
of layer node affect the robustness of multiplex networks?

Aiming to answer this issue,  we consider the robustness of
multiplex networks under layer node-based attack, which can be
further divided into random and targeted scenarios. With the
framework of generating function method \cite{Watts01PRE}, we
propose theoretical method to calculate the critical threshold of
network crash and the size of giant component when a fraction of
layer nodes are removed.  Furthermore, we also
compare the robustness of multiplex networks under multiplex
nodes-based attack and layer node-based attack.

\section{Model and analysis}

As mentioned in previous literatures
\cite{Buldyrev10,Gao11PRL,Parshani10}, the robustness of networks is
usually evaluated by one critical threshold value and the size of
giant component after the removal of nodes. If the fraction of
removed nodes exceeds this critical threshold, the giant component
becomes null. Here it is worth mentioning that the component of multiplex network is defined as a set of connected multiplex nodes. A pair of multiplex nodes is regarded to have connection if there exists at least one type of link between them.
Therefore, attacking some layer nodes may not
destroy their connection with other nodes (see Fig.2). In the following, we will focus on theoretical method of calculating the
critical threshold value and the size of giant component of the
multiplex networks under layer node-based attack.

\begin{figure}[!htb]
\centering
\includegraphics[scale=0.8,trim=50 0 50 0]{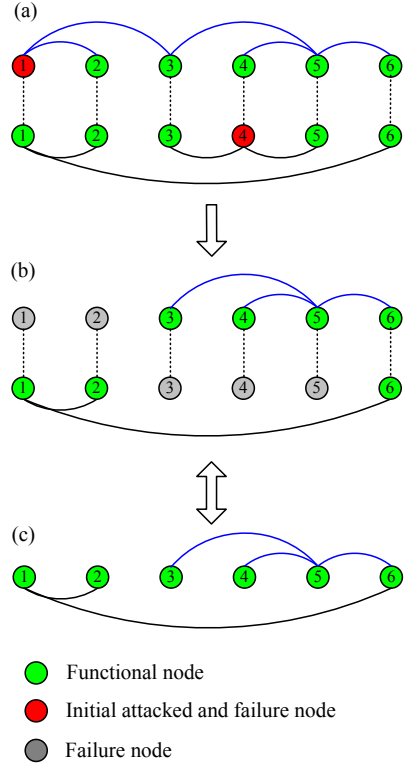}
\caption{Layer node-based attack: (a) layer node 1 of Layer-1 and layer node 4 of Layer-2 are initially attacked, (b) soon layer nodes 1,2 of Layer-1 and 3,4 and 5 of Layer-2 become failure nodes since they do not belong to the giant component of corresponding layers. (c) But multiplex nodes 1-6 still
belong to the giant component of the multiplex networks
since they connect to the giant component through at least one type of links. }
\end{figure}

For a multiplex network composing of $N$ multiplex nodes and $m$
network layers, the generating function for the joint degree
distribution $p(\overrightarrow{k_j})$, where
$\overrightarrow{k_j}=({k_j}_1,{k_j}_2,...,{k_j}_m)$ denotes the
degrees of a multiplex node $j$ in each layer, can be written in the
form of a finite polynomial
\begin{equation}
G_0(\overrightarrow{x})=\sum\limits_{\overrightarrow{k_j}}
p(\overrightarrow{k_j})\prod\limits_{i=1}^{m}x_i^{{k_j}_i},
\end{equation}
where $\overrightarrow{x}=(x_1,x_2,...,x_m)$ represents the
auxiliary variable coupled to $\overrightarrow{k_j}$. Then the
generating function of remaining joint degree distribution by
following a randomly chosen link of  network layer $i$ is given by
\begin{equation}
G_1^{(i)}(\overrightarrow{x})=\frac{1}{z_i}\frac{\partial}{\partial
x_i}G_0(\overrightarrow{x}),
\end{equation}
where $z_i$ is the average degree of layer $i$.

If $u_i$ ($i=1,2,...,m$) is defined as the probability that a
multiplex node reached by following a random chosen link of network
layer $i$ does not belong to the giant component, it can be derived
by the coupled self-consistency equation
\begin{equation}
u_i=G_1^{(i)}(\overrightarrow{u}),
\end{equation}
where $\overrightarrow{u}=(u_1,u_2,...,u_m)$. Furthermore, the size
of the giant component can be calculated according to
\begin{equation}
R=1-G_0(\overrightarrow{u}).
\end{equation}

Along this framework, we can now turn to the layer node-based attack
on multiplex networks. If $\phi_i({k_j}_i)$ is
used to denote the probability that a layer node with degree
${k_j}_i$ is removed from network layer $i$, then the generating
function of the joint degree distribution after the removal of layer
nodes can be expressed as
\begin{equation}
H_0(\overrightarrow{x})=\sum\limits_{\overrightarrow{k_j}}
p(\overrightarrow{k_j})\prod\limits_{i=1}^{m}(\phi_i({k_j}_i)+(1-\phi_i({k_j}_i))x_i^{{k_j}_i}).
\end{equation}
Correspondingly, the generating function of remaining joint degree
distribution after the removal of layer nodes by following a
randomly chosen link of network layer $i$ is given by
\begin{equation}
H_1^{(i)}(\overrightarrow{x})=\frac{1}{z_i}\frac{\partial}{\partial
x_i}H_0(\overrightarrow{x}).
\end{equation}

In the case of layer node removal, the probability $v_i$ that a
multiplex node reached by following one random chosen link of
network layer $i$ does not belong to the giant component can be
written as
\begin{equation}
\begin{split}
& v_i=\frac{1}{z_i}\sum\limits_{\overrightarrow{k_j}}
{k_j}_ip(\overrightarrow{k_j})(\phi_i({k_j}_i)+ \\
&\ \ \ \ \ (1-\phi_i({k_j}_i))v_i^{{k_j}_i-1}\prod\limits_{s\neq
 i}(\phi_s({k_j}_s)+(1-\phi_s({k_j}_s))v_s^{{k_j}_s}))
 \\
&\ \ \ =\frac{\langle {k_j}_i\phi_i({k_j}_i)\rangle
}{z_i}+H_1^{(i)}(\overrightarrow{v}).
 \end{split}
\end{equation}
Then, after the removal of nodes from layers, the size of
giant component is given as follows
\begin{equation}
R=1-H_0(\overrightarrow{v}).
\end{equation}

The existence of giant component under layer node-based attack
requires the largest eigenvalue $\Lambda$ of the Jacobian matrix
$\textbf{J}$ of Eq. (7) at (1,1,...,1) to be larger than unity
\cite{Min14PRE}. In this work, we mainly focus on multiplex networks composed of two Erd{\"o}s-R{\'e}nyi (ER) random \cite{er} or  Barab{\'a}si-Albert scale-free (SF) \cite{ba} network layers (namely, $m=2$), $\textbf{J}$ thus can be written as
\begin{equation}
\emph{\emph{\textbf{J}}}=\left (
  \begin {array}{cc}
  \kappa_1 & \mathcal {K}_1\\
  \mathcal {K}_2 & \kappa_2
  \end{array} \right),
\end{equation}
where
$\kappa_i=(\langle {k_j}_i^2(1-\phi_i({k_j}_i))\rangle -\langle
{k_j}_i(1-\phi_i({k_j}_i))\rangle)/z_i$ and $\mathcal
{K}_i=\langle
{k_j}_1{k_j}_2(1-\phi_1({k_j}_1))(1-\phi_2({k_j}_2))\rangle /z_i.$
The largest eigenvalue $\Lambda$ is given by
\begin{equation}
\Lambda=\frac{1}{2}[\kappa_1+\kappa_2+ \sqrt{\mathstrut
(\kappa_1-\kappa_2)^2+4\mathcal {K}_1\mathcal {K}_2}].
\end{equation}

\section{Results}

\subsection{Layer node-based random attack}

For layer node-based random attack, which is characterized by random
removal of layer nodes from network layers, there exists the removal
probability $\phi_i({k_j}_i)=\phi_i^{LR}\ (i=1,2;\ j=1,2,...N)$.
According to the above analysis, the critical threshold and the size
of giant component of multiplex networks under layer node-based
random removal can be respectively expressed as
\begin{equation}
(\phi_1^{LR},\phi_2^{LR})_c=\{(\phi_1^{LR},\phi_2^{LR})|\Lambda=1\}
\end{equation}
and
\begin{equation}
R^{LR}=1-H_0(\overrightarrow{v}),
\end{equation} where $\phi_i({k_j}_i)=\phi^{LR}_i$.

It is worth mentioning that above $(\phi_1^{LR},\phi_2^{LR})_c$
there is no giant component, whereas below
$(\phi_1^{LR},\phi_2^{LR})_c$ a giant connected cluster exists.

\begin{figure}[!htb]
\centering
\includegraphics[scale=0.4,trim=50 0 50 0]{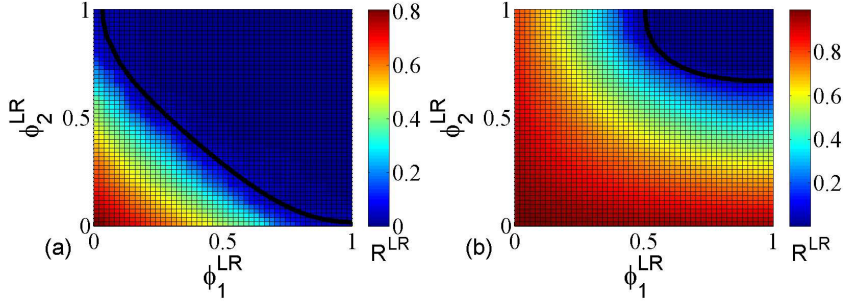}
\caption{(Color online) The size $R^{LR}$ of giant component in dependence on removal probability $\phi_1^{LR}$ and $\phi_2^{LR}$ for layer node-based random attack.  The black line indicates the theoretical critical threshold calculated according to Eq.(11). The networks used are multiplex ER network with average degree (a) $z_1=z_2=1$, (b) $z_1=2,\ z_2=3$ and size $N=5000$.}
\end{figure}

\begin{figure}[!htb]
\centering
\includegraphics[scale=0.4,trim=50 0 50 0]{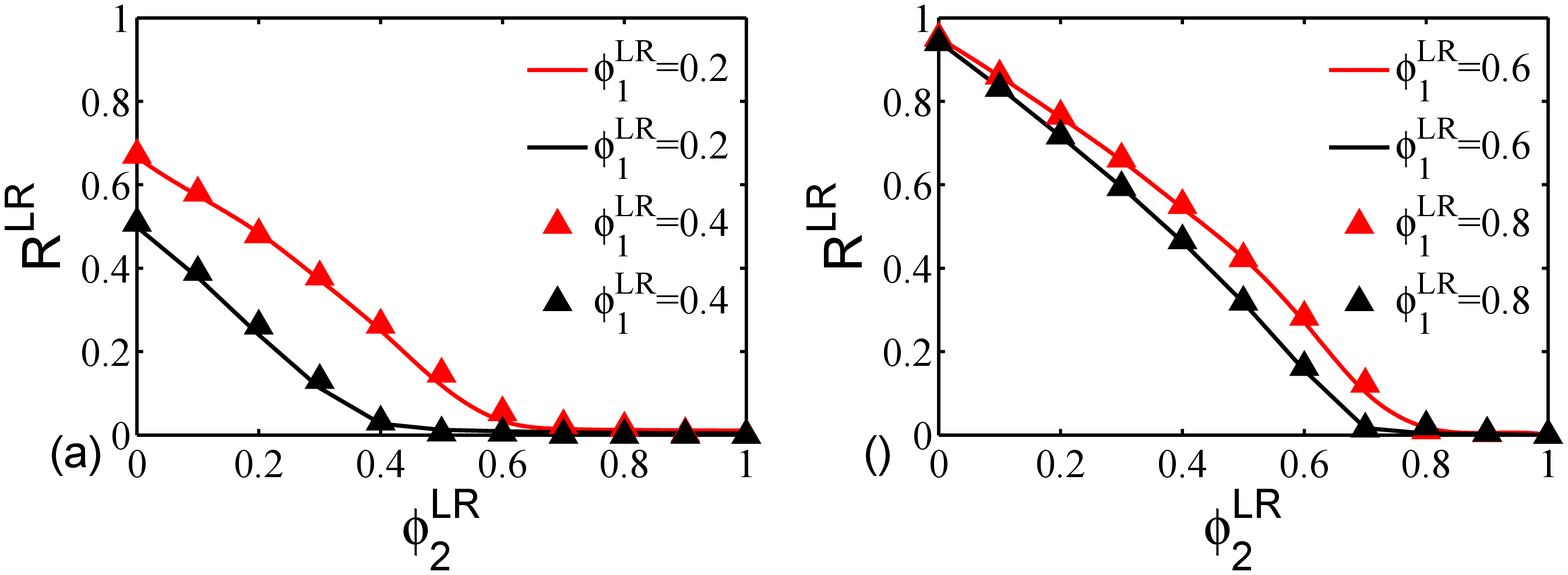}
\caption{(Color online) Theoretical (line) and numerical (point)
results of the size $R^{LR}$ of giant component as a function of
$\phi_2^{LR}$ when $\phi_1^{LR}$ takes fixed values. The networks
used are multiplex ER networks with average degree (a) $z_1=z_2=1$,
(b) $z_1=2,\ z_2=3$ and size $N=5000$.}
\end{figure}

We start by inspecting how layer node-based random attack affects
the robustness of multiplex networks. Fig.3 shows the size $R^{LR}$
of giant component in dependence on the removal probability
$\phi_1^{LR}$ and $\phi_2^{LR}$ for network layer 1 and network
layer 2, respectively. Moreover, the black line indicates the
theoretical critical threshold calculated according to Eq.(11). It
is clear that  when the removal probability ($\phi_1^{LR},
\phi_2^{LR}$) is above this black line, the size of
giant component becomes negligible; whereas there exists one giant
component if ($\phi_1^{LR}, \phi_2^{LR}$) is located below this
black line. This implies that the theoretical critical threshold can
accurately predict the impact of layer node-based attack on
robustness of multiplex networks. To further validate this fact, we
also compare the theoretical prediction derived from Eq.(12) and
simulation results for the size of giant component in Fig.4. It can
be observed that there is indeed good agreement between simulation
and theoretical prediction.

\subsection{Layer node-based targeted attack}

Targeted attack, as a well-known attack strategy, usually aims to
remove influential nodes, which can be identified by centrality measures, such as
the degree centrality, eigenvector centrality, $k$-shell centrality
and betweenness centrality \cite{Ren14CSB}. In this work, we mainly pay attention to the viewpoint of degree centrality. For layer node-based targeted
attack, the removal probability of a layer node with degree
${k_j}_i$ is determined by its degree, and can be expressed as
follows
\begin{equation}\phi_i({k_j}_i)=\left\{\begin{array}{ll} 1,\ \ \ \emph{\emph{if}}\ {k_j}_i>{k_c}_i\\
f_i,\ \ \ \emph{\emph{if}}\ {k_j}_i={k_c}_i\\ 0,\ \ \
\emph{\emph{if}}\ {k_j}_i<{k_c}_i\end{array},\right.\end{equation}
where ${k_c}_i$ is the cutoff degree for attack on network layer
$i$, and $f_i$ denotes the removal probability of node with degree
${k_c}_i$. Consequently, the total fraction of removal nodes in
network layer $i$ is given by
\begin{equation}
\phi_i^{LT}=\sum\limits_{{k_j}_i}p_i({k_j}_i)\phi_i({k_j}_i),
\end{equation}
where $p_i({k_j}_i)$ indicates the fraction of layer nodes with
degree ${k_j}_i$ in layer $i$.

Similar to Eqs. (11) and (12), we can get the critical threshold
\begin{equation}
\begin{split}
& (\phi_1^{LT},\phi_2^{LT})_c=\{(\phi_1^{LT},\phi_2^{LT})|\Lambda=1\},
\end{split}
\end{equation}
and the size of giant component
\begin{equation}
R^{LT}=1-H_0(\overrightarrow{v}),
\end{equation} where $\phi_i({k_j}_i)$ is defined as Eq. (13),
for layer node-based targeted attack on multiplex networks consisting of two network layers.

\begin{figure}[!htb]
\centering
\includegraphics[scale=0.4,trim=50 0 50 0]{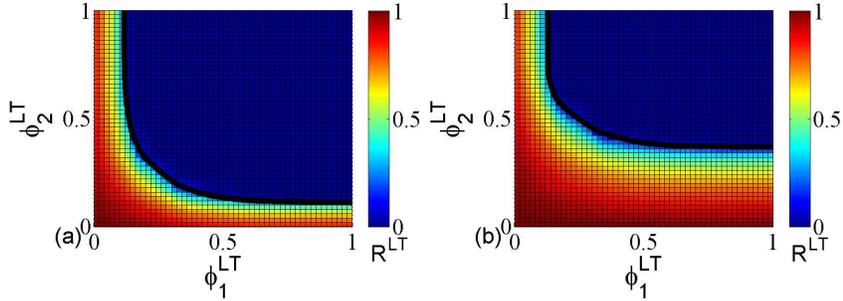}
\caption{(Color online) The size $R^{LT}$ of giant component in
dependence on removal probability $\phi_1^{LT}$ and $\phi_2^{LT}$
for layer node-based targeted attack. The black line indicates the
theoretical critical threshold calculated according to Eq.(15). The
networks used are multiplex ER networks with average degree (a)
$z_1=z_2=2$, (b) $z_1=2,\ z_2=4$ and size $N=5000$.}
\end{figure}

\begin{figure}[!htb]
\centering
\includegraphics[scale=0.4,trim=50 0 50 0]{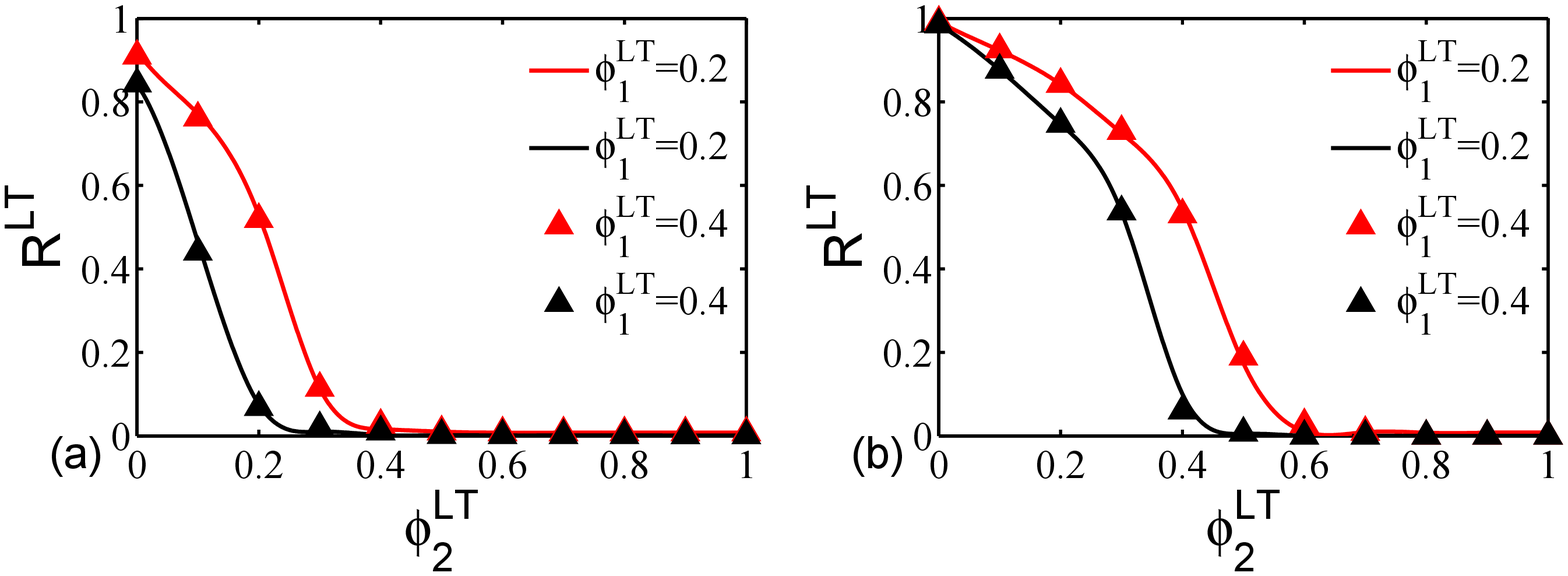}
\caption{(Color online) Theoretical (line) and numerical (point)
results of the size $R^{LT}$ of giant component as a function of
$\phi_2^{LT}$ when $\phi_1^{LT}$ takes fixed values. The networks
used are multiplex ER networks with average degree (a) $z_1=z_2=2$,
(b) $z_1=2,\ z_2=4$  and size $N=5000$.}
\end{figure}

In Fig.5, the color code represents the size $R^{LT}$ of the giant
component as a function of the removal probability $\phi_1^{LT}$ and
$\phi_2^{LT}$ under layer node-based targeted attack, and the black
line indicates the theoretical critical threshold calculated
according to Eq.(15). Similar to Fig.3, the theoretical prediction
fully agrees with the simulation results. Moreover, Fig.6 provides
the further comparison between the theoretical prediction and
simulation for  the size of giant components, which also validates
the accuracy of theoretical method. Combining with all the above
phenomena, it is clear that the proposed theoretical framework can
allow us to accurately calculate the critical threshold and the size of giant component under the layer node-based attack.

\subsection{Comparison of robustness of multiplex networks}

Based on the above analysis, multiplex node-based attack proposed in
\cite{Min14PRE}, can be regarded as a special case of layer
node-based attack when all the removed nodes or replicas are the
same in each network layer. From the economic viewpoint, the cost of
removing $p$ fraction of  multiplex nodes seems approximately equal
to that of removing $p$ fraction of layer nodes in each network
layer. However, the damage of both scenarios on the multiplex
networks may be greatly different. In this sense, it becomes very
instructive to compare the robustness of multiplex networks under
multiplex node-based attack and layer node-based attack. For
simplicity of comparison, we assume that layer node-based attack means to remove
the same proportion of layer nodes in each network layer in what
follows. The removal probability correspondingly becomes
$\phi_1^{LR}=\phi_2^{LR}=\phi^{LR}$ for layer node-based random
attack and $\phi_1^{LT}=\phi_2^{LT}=\phi^{LT}$ for layer node-based
targeted attack. While for multiplex node-based attack, the total fraction of removal multiplex nodes under random
attack and targeted attacks becomes $\phi^{MR}$ (all of the multiplex nodes are removed randomly with probability $\phi^{MR}$) and
\begin{equation}
\phi^{MT}=\sum\limits_{\overrightarrow{k_j}}p(\overrightarrow{k_j})\phi^{MT}(\overrightarrow{k_j}),
\end{equation}
where $p(\overrightarrow{k_j})$ indicates the fraction of multiplex nodes with
degree $\overrightarrow{k_j}=\{k_{j_1},k_{j_2}\}$, and $\phi^{MT}(\overrightarrow{k_j})$ is defined as the removal probability of multiplex nodes with
degree $\overrightarrow{k_j}$ and given by \begin{equation}\phi^{MT}(\overrightarrow{k_j})=\left\{\begin{array}{ll} 1,\ \ \ \emph{\emph{if}}\ k_{j_1}+k_{j_2}>{k_c}\\
f,\ \ \ \emph{\emph{if}}\ k_{j_1}+k_{j_2}={k_c}\\ 0,\ \ \
\emph{\emph{if}}\ k_{j_1}+k_{j_2}<{k_c}\end{array},\right.\end{equation}
where ${k_c}$ is the cutoff degree and $f$ denotes the removal probability of node which satisfies $k_{j_1}+k_{j_2}={k_c}$.

\begin{figure}[!htb]
\centering
\includegraphics[scale=0.4,trim=50 0 50 0]{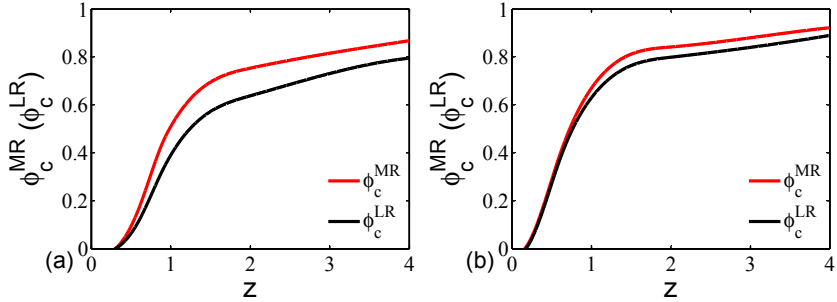}
\caption{(Color online) The critical threshold of multiplex networks
in dependence on the network average degree under multiplex
node-based random attack (red dash line) and layer node-based random
attack (black solid line). The networks used are (a) multiplex ER
networks with average degree $z_1=z_2=z$ and (b) multiplex SF
networks with average degree $z_1=z_2=z$. The size of all the
networks is $N=5000$.}
\end{figure}

\begin{figure}[!htb]
\centering
\includegraphics[scale=0.4,trim=50 0 50 0]{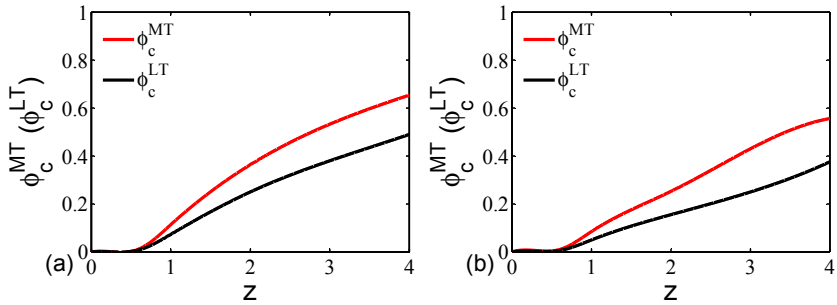}
\caption{(Color online) The critical threshold of multiplex networks
in dependence on the network average degree under multiplex
node-based targeted attack (red line) and layer node-based
targeted attack (black line). The networks used are (a)
multiplex ER networks with average degree $z_1=z_2=z$ and (b)
multiplex SF networks with average degree $z_1=z_2=z$. The size of
all the networks is $N=5000$.}
\end{figure}

Similar to the above treatment, we still use the critical threshold
as a uniform evaluation index for multiplex node-based attack and
layer node-based attack. In fact, the larger the value of  critical
threshold, the better the robustness of multiplex networks against
attack. Fig.7 features how the the critical threshold of multiplex
networks varies as a function of average degree under both multiplex
node-based random attack (red line) and layer node-based random
attack (black line). It is clear that the threshold of both
cases rises with the increment of average degree, which means that
multiplex networks are more robust for denser connections.
Interestingly, another observation of utmost significance is that
the threshold of multiplex node-based random attack is always higher
than that of layer node-based random attack, irrespective of the
average degree and underlying connection topology. This is to say,
multiplex networks are more vulnerable under layer node-based
attack, because it usually makes more multiplex nodes subject to
attack and lose more connections with other multiplex nodes.
Moreover, we can also obtain the similar observation for multiplex
node-based targeted attack and layer node-based targeted attack in
Fig.8, which further supports the fact that layer node-based attack
brings larger damage to multiplex networks. Along this seminal
finding, it may shed new light into the research of protection or
immunization of empirical multiplex topology.

\section{Summary}

To sum, we have studied the robustness of multiplex networks under
layer node-based attack. Under this framework, the layer nodes can
be removed randomly or intentionally, which corresponds to layer
node-based random attack or layer node-based targeted attack.  A
theoretical method is proposed to evaluate the robustness of
multiplex networks when a fraction of layer nodes are removed.
Through numerous simulations, this method can accurately calculate the threshold and size of giant component, irrespective of the removal
case. In addition, we also compare the robustness of multiplex
networks under multiplex node-based attack and layer node-based
attack. An interesting finding is that multiplex networks will be
more robust under multiplex node-based attack, which is
universal for different average degree and underlying topology. With
regard to the reason, it may be related with the fact that layer
node-based attack usually brings damage to more multiplex nodes,
which will directly break the remaining joint component of networks.

Since multiplex framework is ubiquitous in realistic social and
technological networks, we hope that the present outcomes can
inspire further research of the robustness of multiplex networks,
especially combining with the novel properties of multiplex
networks, like the clustering characteristic \cite{Parshani10EPL},
degree-degree correlation between network layers \cite{Zhao14PLA}.
In addition, the targeted attack can also be incorporated into other
centrality measures, such as the eigenvector centrality, $k$-shell
centrality and betweenness centrality \cite{Ren14CSB}. Along this
line, we may get new understanding for the protection of multiplex
network.

\section{Acknowledgement}
This paper was supported by the National Natural Science Foundation of China (Grant No. 61572297), Shandong Province Outstanding Young Scientists Research Award Fund Project (Grant No. BS2015DX006, BS2014DX007) and Natural Science Foundation of Shandong Province (Grant No. ZR2014FM003, ZR2015YL018).

\end{document}